\documentclass[prd]{revtex4}
\usepackage{graphicx}
\usepackage{dcolumn}
\usepackage{bm}

\begin{document}

\preprint{YITP-04-74,gr-qc/0412070}

\title{Scalar field perturbation on six-dimensional ultra-spinning
black holes}

\author{Yoshiyuki Morisawa}
 \email{morisawa@yukawa.kyoto-u.ac.jp}
\affiliation{%
Yukawa Institute for Theoretical Physics, Kyoto University,
Kyoto 606-8502, Japan
}%

\author{Daisuke Ida}
 \email{daisuke.ida@gakushuin.ac.jp}
\affiliation{%
Department of Physics, Gakushuin University,
Tokyo 171-8588, Japan
}%

\date{\today}

\begin{abstract}
We have studied the scalar field perturbations on
six-dimensional ultra-spinning black holes.
We have numerically calculated the quasinormal modes of rotating black holes.
Our results suggest that such perturbations are stable.
\end{abstract}


\maketitle

\section{Introduction}
\label{sec:introduction}

The production of higher-dimensional black holes
in collider~\cite{Banks:1999gd,Giddings:2001bu,Dimopoulos:2001hw}
is the important prediction arising from 
the large extra dimension scenario~\cite{Arkani-Hamed:1998rs,Antoniadis:1998ig}
or the warped compactification (RS1) scenario~\cite{Randall:1999ee},
where the fundamental (higher-dimensional) scale of gravitation can be set
to the order of TeV.
Such phenomena would strongly suggest the presence of the extra dimensions,
and we would also obtain the experimental method to seek for the quantum gravity.
Hence, it is important to investigate the fundamental properties of
higher-dimensional black holes such as stability, which is the subject of the 
present paper.

There is a variety of the higher-dimensional black objects 
with nontrivial topology
unlike those in four-dimensional space-times
where each component of stationary black hole is a topological two-sphere.
In fact, higher-dimensional vacuum Einstein equation admits
simple solutions called black branes which are direct product of the
four-dimensional Schwarzschild black hole and the flat space.
We are usually interested in the case where the flat dimensions 
in the black brane space-time are compactified into say a torus.

Gregory and Laflamme~\cite{Gregory:1993vy} have found that the black
branes are unstable under linear gravitational perturbations
when the compactification scales are large compared with the Schwarzschild radius.
On the other hand,  Ishibashi and Kodama~\cite{Ishibashi:2003ap} have shown that the
higher-dimensional Schwarzschild black holes
in the spherically symmetric space-times
are gravitationally stable as in the case with the four space-time dimensions.
An accurate way to compute the quasinormal frequencies of
higher-dimensional Schwarzschild black holes has been obtained via 6th
order WKB approximation~\cite{Konoplya:2003ii}.

It is also important to consider the stability of the rotating black
holes, since black holes produced in the collider generically 
have angular momenta and the production cross section is larger for
more rapidly rotating black holes~\cite{Ida:2002ez}
In contrast to the Kerr-bound on four-dimensional rotating black
holes, there are black hole solutions with an arbitrary large angular
momentum for a fixed mass in higher-dimensional general relativity.
Recently, the existence of the effective Kerr-bound on such rapidly
rotating black holes in higher dimension has been conjectured by
Emparan and Myers~\cite{Emparan:2003sy}.
They showed that the geometry of the event horizon of such rapidly
rotating black holes in six or higher dimension behaves like the black
membranes and argued that such rapidly rotating black holes therefore become unstable.

Although we should analyze the gravitational perturbations of 
rotating black holes to confirm such instability, 
any formalism where the perturbative equations are
separable is yet unknown.
Here we just consider the perturbation of the massless scalar field
propagating on the rotating black hole space-times.
Even if we restrict ourselves to field perturbation on fixed
background space-time, there exists an instability of massless field
perturbation on rotating black branes and strings discovered by
Cardoso and Lemos~\cite{Cardoso:2004zz}.
The equation of motion for the massless scalar field perturbation on
the rotating black holes is known to be separable in any
dimension~\cite{Ida:2002zk,Frolov:2002xf}, and authors examined the
stability of this system in five-dimension as the simplest case in the
previous work~\cite{Ida:2002zk}.
The present paper is the complementary work on the renewed interest in 
the instability (or stability) of the rapidly rotating black holes in
six dimensions.

\section{Myers-Perry black hole and its membrane limit}
\label{sec:MyersPerry}

The $(4+n)$-dimensional Myers-Perry metric~\cite{Myers:1986un} with
only one nonzero angular momentum is given by
\begin{eqnarray}
\label{eq:MyersPerry}
g &=&
-{\Delta-a^2\sin^2\vartheta \over \Sigma}dt^2
-{2a(r^2+a^2-\Delta) \over \Sigma}\sin^2\vartheta dt d\varphi
\nonumber\\&&
+{(r^2+a^2)^2-\Delta a^2\sin^2\vartheta \over \Sigma}\sin^2\vartheta d\varphi^2
+{\Sigma \over \Delta}dr^2
+\Sigma d\vartheta^2
+r^2\cos^2\vartheta d\Omega_n^2,
\end{eqnarray}
where
\begin{eqnarray}
\Sigma &=& r^2+a^2\cos^2\vartheta,
\\
\Delta &=& r^2+a^2-\mu r^{1-n},
\end{eqnarray}
and $d\Omega_n^2$ denotes the standard metric of the unit $n$-sphere.
This metric describes a rotating black hole in asymptotically flat,
vacuum space-time with the mass and angular momentum proportional to
$\mu$ and $\mu a$, respectively.
Hereafter, $\mu, a>0$ are assumed.
The event horizon is located at $r=r_H$, such that
$\Delta|_{r=r_H}=0$, which is homeomorphic to $S^{n+2}$.
For $n\geq 2$, $\Delta$ always has a positive root for arbitrary $a$,
namely regular black hole solutions exist with arbitrarily large
angular momentum per unit mass.

In the limit where $a\rightarrow\infty$, the coordinate radius of the
horizon is approximated by $r_H\approx (\mu/a^2)^{1/(n-1)}$.
Hence, the horizon radius $r_H$ shrinks to zero when one take the
limit where $a\rightarrow\infty$ with fixed $\mu$.
To avoid a vanishing horizon radius, Emparan and Myers introduced the black
membrane limit where $a\rightarrow\infty$ and $\mu\rightarrow\infty$
with fixed $\hat{\mu}:=\mu/a^2$.
In this limit, the Myers-Perry metric behaves like a black membrane 
metric near the pole $\vartheta=0$:
\begin{eqnarray}
g &=&
-\left(1-{\hat{\mu}\over r^{n-1}}\right)dt^2
+{dr^2 \over 1-\hat{\mu}/r^{n-1}}
+r^2 d\Omega_n^2 + d\varsigma^2 + \varsigma^2 d\varphi^2,
\end{eqnarray}
where the new coordinate is defined by $\varsigma := a\sin\vartheta$.
In this paper, we are concerned with the dimensionless Kerr
parameter $a_*:=a/r_H$.
The limit where $a_* \rightarrow \infty$ with fixed $r_H$ is
consistent to the membrane limit.

\section{Scalar field equations}
\label{sec:perturbation}

Here we consider the perturbation of the massless scalar field on the
background Myers-Perry metric~(\ref{eq:MyersPerry}).
We put the scalar field configuration as
\begin{eqnarray}
\phi &=&
e^{i \omega t-i m \varphi}R(r)S(\vartheta)Y(\Omega),
\end{eqnarray}
where $Y(\Omega)$ is the hyperspherical harmonics on the $n$-sphere
with the eigenvalue $j(j+n-1)$, $(j=0,1,2,\cdots)$.
Then we obtain the field equations with separated variables:
\begin{eqnarray}
\label{eq:angular}
{1 \over \sin\vartheta\cos^n\vartheta}
\left({d\over d\vartheta}\sin\vartheta\cos^n
\vartheta{dS\over d\vartheta}\right)
+[\omega^2a^2\cos^2\vartheta-m^2\csc^2\vartheta
-j(j+n-1)\sec^2\vartheta+A]S &=& 0,
\end{eqnarray}
and
\begin{eqnarray}
\label{eq:radial}
{1\over r^n}{d\over dr}\left(r^n\Delta{dR\over dr}\right)
+\left\{{[\omega(r^2+a^2)-ma]^2 \over \Delta}
-{j(j+n-1)a^2 \over r^2}-\lambda\right\}R &=& 0,
\end{eqnarray}
where $\lambda:=A-2m\omega a+\omega^2a^2$.

We will solve the eigenvalue problem under the quasinormal boundary
condition:
\begin{eqnarray}
R\sim \left\{
{(r-r_H)^{i\sigma} \quad (r\rightarrow r_H),
\atop
r^{-(n+2)/2} e^{-i\omega r} \quad (r\rightarrow +\infty),}
\right.
\end{eqnarray}
where
\begin{eqnarray}
\sigma:=
{(r_H^2+a^2)\omega-m a \over (n-1)(r_H^2+a^2)+2r_H^2}r_H.
\end{eqnarray}

When $a=0$, the eigenfunctions for the angular
equation~(\ref{eq:angular}) are analytically given in terms of the
hypergeometric functions:
\begin{eqnarray}
S_{\ell j m} &=&
(\sin\vartheta)^{|m|}(\cos\vartheta)^j
F\left(-\ell,\ell+j+|m|+{n+1\over 2},j+{n+1\over 2};
\cos^2\vartheta\right),
\end{eqnarray}
with the eigenvalues
\begin{eqnarray}
A_{\ell j m} = (j+|m|+2\ell)(j+|m|+2\ell+n+1),
\quad (\ell=0,1,2,\cdots).
\end{eqnarray}
Then $e^{-im\varphi}S_{\ell j m}(\vartheta)Y(\Omega)$ behaves as the
hyperspherical harmonics on $S^{n+2}$, which belongs to the
irreducible representation of the rotational group $SO(n+3)$.
Especially for $n=2$, the hyperspherical function on $S^4$ with the
eigenvalue $L(L+3)$ belongs to the representation characterized by the
Dynkin index $[L,0]$ and has $(L+1)(L+2)(2L+3)/6$-fold degeneracy.
When $a\ne 0$, $SO(5)$ symmetry of the space-time breaks down to
$SO(2)\times SO(3)$ and the degenerate mode partially split into the
several modes with $(2j+1)$-fold degeneracies, where $j$ and $m$ are
constrained by $L=j+|m|+2\ell$, $(\ell=0,1,2,\cdots)$.

\section{Numerical computation}
\label{sec:numerical}

We use the continued fraction method to determine the resonant
frequency $\omega$ and the separation constant $A$.
We assume the following series expansion for $S$:
\begin{eqnarray}
S=(\sin\vartheta)^{|m|}(\cos\vartheta)^j
\sum_{k=0}^{\infty} a_k (\cos^2\vartheta)^k,
\end{eqnarray}
which automatically satisfies the regular boundary conditions at
$\vartheta=0,\pi/2$ and converges at any time.
Substituting this expansion into the angular
equation~(\ref{eq:angular}), we obtain the three-term recurrence
relations:
\begin{eqnarray}
\label{eq:angrec}
\bar{\alpha}_ka_k+\bar{\beta}_ka_{k-1}+\bar{\gamma}_ka_{k-2}
&=& 0, \, (k=1,2,\cdots),
\end{eqnarray}
where $a_{-1}=0$ and the coefficients are given by
\begin{eqnarray}
\bar{\alpha}_k &=& -8k(2j+2k+n-1),
\\
\bar{\beta}_k &=& 2[(j+|m|+2k-2)(j+|m|+2k+n-1)-A],
\\
\bar{\gamma}_k &=& -a_*^2\omega_*^2.
\end{eqnarray}
Here $a_*:=a/r_H$ and $\omega_*:=\omega r_H$ are the dimensionless
quantities.
The limit where $a_*\rightarrow\infty$ corresponds to the membrane
limit as mentioned above.

We expand the radial function $R$ as
\begin{eqnarray}
\label{eq:radexp}
R &=&
e^{-i\omega_*r/r_H}\left({r-r_H \over r_H}\right)^{i\sigma}
\left({r+r_H \over r_H}\right)^{-(n+2)/2-i\sigma}
\sum_{k=0}^{\infty}b_k\left({r-r_H\over r+r_H}\right)^k,
\end{eqnarray}
which automatically satisfies the quasinormal boundary conditions
 and converges at any time.
When $n=2$, substituting Eq.~(\ref{eq:radexp}) into the radial
equation~(\ref{eq:radial}), we obtain the eight-term recurrence
relations
\begin{eqnarray}
\label{eq:eight-term}
\alpha_kb_k+\beta_kb_{k-1}+\gamma_kb_{k-2}+\delta_kb_{k-3}
+\epsilon_kb_{k-4}+\zeta_kb_{k-5}+\eta_kb_{k-6}+\theta_kb_{k-7}
&=& 0, \, (k=1,2,\cdots),
\end{eqnarray}
where $b_{-1}=b_{-2}=\cdots=0$ and the coefficients are given by
\begin{eqnarray}
\alpha_{k} &=&
-(a_*^2+1) (a_*^2+3)^2 k (k+2i\sigma),
\\
\beta_{k} &=&
(a_*^2+3) \{
-2 (5 a_*^4+22 a_*^2+33) \sigma^2
-4 m a_* (a_*^2+5) \sigma,
\nonumber\\&&
+(a_*^2+1)
[(5 k^2-7 k+4) a_*^2+3 k^2-3 k+6 + 2 j (j+1) a_*^2 + 2 \lambda] \}
\nonumber\\&&
+i (a_*^2+3) \{
[(14 k-9) a_*^4+(40 k-22) a_*^2+42 k-21] \sigma
+ 2 m a_* (a_*^2+3) (2 k-1) \},
\\
\gamma_{k} &=&
(27 a_*^6+117 a_*^4+89 a_*^2-129) \sigma^2
+4 m a_* (a_*^2+3) (5 a_*^2+1) \sigma
+4 m^2 a_*^2 (a_*^2+5)
\nonumber\\&&
-(a_*^2+1)
[(9 k^2-26 k+24) a_*^4+(14 k^2-42 k+40) a_*^2-3 k^2-24
+4 j (j+1) a_*^2 (2 a_*^2+3)
-12 \lambda]
\nonumber\\&&
-i\{
[(30 k-42) a_*^6+(106 k-152) a_*^4+(58 k-114) a_*^2-114 k+108] \sigma
+4 m a_* (a_*^2+3) [(3 k-4) a_*^2-3 (k-1)]
\},
\nonumber\\&&
\\
\delta_{k} &=&
-(3 a_*^2-1) (5 a_*^4+6 a_*^2-15) \sigma^2
-8 m a_* (2 a_*^2+3) (a_*^2-3) \sigma
-4 m^2 a_*^2 (a_*^2-7)
\nonumber\\&&
+(a_*^2+1)[
(5 k^2-25 k+36) a_*^4-2 (k-1) (3 k-2) a_*^2-3 k^2+3 k+24
+4 j (j+1) a_*^2 (3 a_*^2+2)
-4 (a_*^2-2) \lambda
]
\nonumber\\&&
+i\{
[(18 k-45) a_*^6+(6 k-55) a_*^4+(-42 k+25) a_*^2+66 k-141] \sigma
+4 m a_* [(2 k-5) a_*^4+(-4 k+5) a_*^2+6 k-12]
\},
\nonumber\\&&
\\
\epsilon_{k} &=&
-2 (9 a_*^2+25) (a_*^2+1)^2 \sigma^2
-8 m a_* (2 a_*^4+3 a_*^2-3) \sigma
-4 m^2 a_*^2 (a_*^2-3)
\nonumber\\&&
+(a_*^2+1)
[(5 k^2-20 k+16) a_*^4+(14 k^2-60 k+64) a_*^2+5 k^2-24 k+40
-4 j (j+1) a_*^2 (2 a_*^2+1) + 4 \lambda]
\nonumber\\&&
+i\{
[(18 k-36) a_*^6+(78 k-164) a_*^4+(110 k-252) a_*^2+82 k-204] \sigma
+4 m a_* [(2 k-4) a_*^4+(4 k-9) a_*^2+6 k-15]
\},
\nonumber\\&&
\\
\zeta_{k} &=&
(a_*^2+1) \{
4 (6 a_*^4+15 a_*^2+5) \sigma^2
+20 m a_* (a_*^2+1) \sigma
+4 m^2 a_*^2
\nonumber\\&&
-(k-3) (9 k-28) a_*^4-(10 k^2-60 k+86) a_*^2-(k-2) (5 k-19)
+2 (a_*^2+1) [j (j+1) a_*^2 + \lambda]
\}
\nonumber\\&&
-i (a_*^2+1) \{
[(30 k-93) a_*^4+(52 k-160) a_*^2-2 k+13] \sigma
+2 m a_* [(6 k-19) a_*^2-2 k+7]
\},
\\
\eta_{k} &=&
(a_*^2+1)^2 \{
-(9 a_*^2+13) \sigma^2 -4 m a_* \sigma +(k-4) [(5 k-18) a_*^2+k-4]
\}
\nonumber\\&&
+ 2 i (a_*^2+1)^2 \{ [(7 k-27) a_*^2+7 k-28] \sigma + 2 m a_* (k-4) \},
\\
\theta_{k} &=&
(a_*^2+1)^3 [ \sigma^2 - (k-4) (k-5) -i (2 k-9) \sigma].
\end{eqnarray}
After we repeatedly use the Gaussian elimination as written in the
Appendix, the eight-term recurrence relations (\ref{eq:eight-term})
are reduced to the three-term ones
\begin{eqnarray}
\label{eq:radrec}
\alpha'''''_kb_k+\beta'''''_kb_{k-1}+\gamma'''''_kb_{k-2}=0,
\quad (k=1,2,\cdots).
\end{eqnarray}

For given the three-term recurrence relations~(\ref{eq:angrec}), the 
successive coefficients are obtained in two ways, the finite and
infinite continued-fraction representations: 
\begin{eqnarray}
\label{eq:radcontfrac}
{a_{k} \over a_{k-1}} &=&
-{\bar{\beta}_{k} \over \bar{\alpha}_{k}}
-{1 \over \bar{\alpha}_{k}}
{-\bar{\gamma}_{k}\bar{\alpha}_{k-1} \over \bar{\beta}_{k-1}+}
\cdots
{-\bar{\gamma}_{3}\bar{\alpha}_{2} \over \bar{\beta}_{2}+}
{-\bar{\gamma}_{2}\bar{\alpha}_{1} \over \bar{\beta}_{1}}
\nonumber\\
&=&
-{\bar{\gamma}_{k+1} \over \bar{\beta}_{k+1}+}
{-\bar{\alpha}_{k+1}\bar{\gamma}_{k+2} \over \bar{\beta}_{k+2}+}
{-\bar{\alpha}_{k+2}\bar{\gamma}_{k+3} \over \bar{\beta}_{k+3}+}
\cdots,
\end{eqnarray}
where $(1/x+)(y/z)$ is an abbreviation for $1/[x+(y/z)]$.
Similarly for Eq.~(\ref{eq:radrec}), we obtain
\begin{eqnarray}
\label{eq:angcontfrac}
{b_{k'} \over b_{k'-1}} &=&
-{\beta'''''_{k'} \over \alpha'''''_{k'}}
-{1 \over \alpha'''''_{k'}}
{-\gamma'''''_{k'}\alpha'''''_{k'-1} \over \beta'''''_{k'-1}+}
\cdots
{-\gamma'''''_{3}\alpha'''''_{2} \over \beta'''''_{2}+}
{-\gamma'''''_{2}\alpha'''''_{1} \over \beta'''''_{1}}
\nonumber\\
&=&
-{\gamma'''''_{k'+1} \over \beta'''''_{k'+1}+}
{-\alpha'''''_{k'+1}\gamma'''''_{k'+2} \over \beta'''''_{k'+2}+}
{-\alpha'''''_{k'+2}\gamma'''''_{k'+3} \over \beta'''''_{k'+3}+}
\cdots.
\end{eqnarray}
For fixed $k$ and $k'$, we obtain nonlinear algebraic
equations~(\ref{eq:radcontfrac}) and (\ref{eq:angcontfrac}) for two
unknown complex numbers $\omega_*$ and $A$.
We use a standard nonlinear root search algorithm provided by the
{\sc MINPACK} subroutine package to evaluate the numerical solutions
of these equations.
The equations for different sets of $(k,k')$ are used as numerical
check.

When the perturbation become unstable under the membrane limit, we
expect that the behavior of such instability resembles one of the
Gregory-Laflamme instability.
Since they found that the $s$ wave is unstable, we consider the $j=0$ modes.
The critical value of the wavelength of the Gregory-Laflamme
instability is of the order of $r_H$.
In the coordinate $\varsigma$ along the rotation plane,
the perturbation has the wave length
$\varsigma \sim a\vartheta \sim a_*r_H/\ell$ for $\ell$-th mode.
Hence, we expect that the possible instabilities occur for $a_* > \ell$.

\begin{figure}
\rotatebox{270}{\resizebox{6cm}{!}{\includegraphics{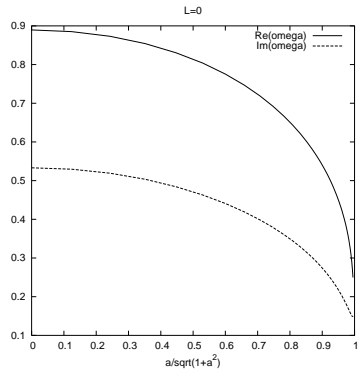}}}
\caption{\label{fig:L0}The quasinormal modes for $(j,m,\ell)=(0,0,0)$.}
\end{figure}

\begin{figure}
\rotatebox{270}{\resizebox{6cm}{!}{\includegraphics{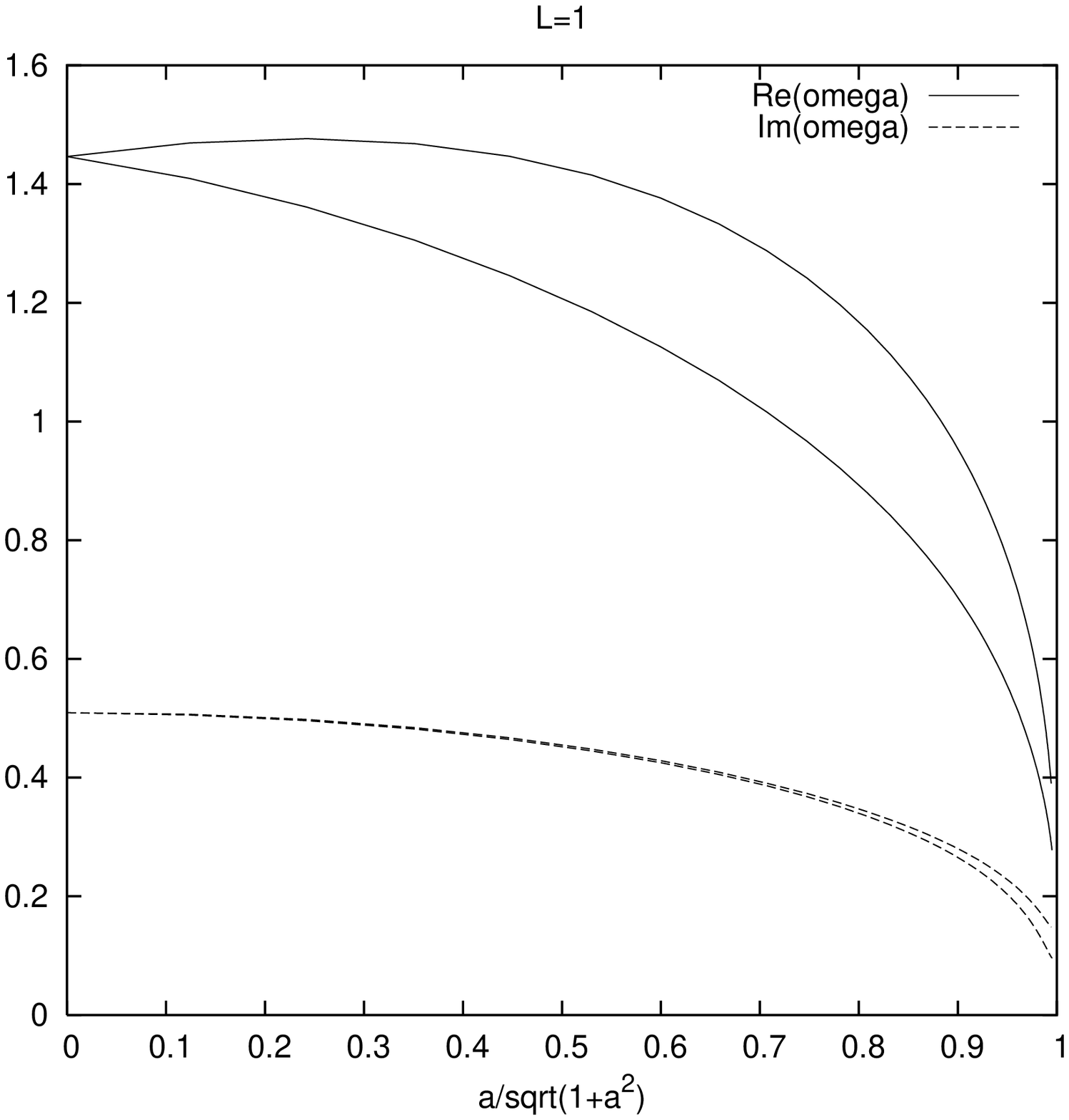}}}
\caption{\label{fig:L1}The quasinormal modes for $(j,m,\ell)=(0,\pm1,0)$.}
\end{figure}

\begin{figure}
\rotatebox{270}{\resizebox{6cm}{!}{\includegraphics{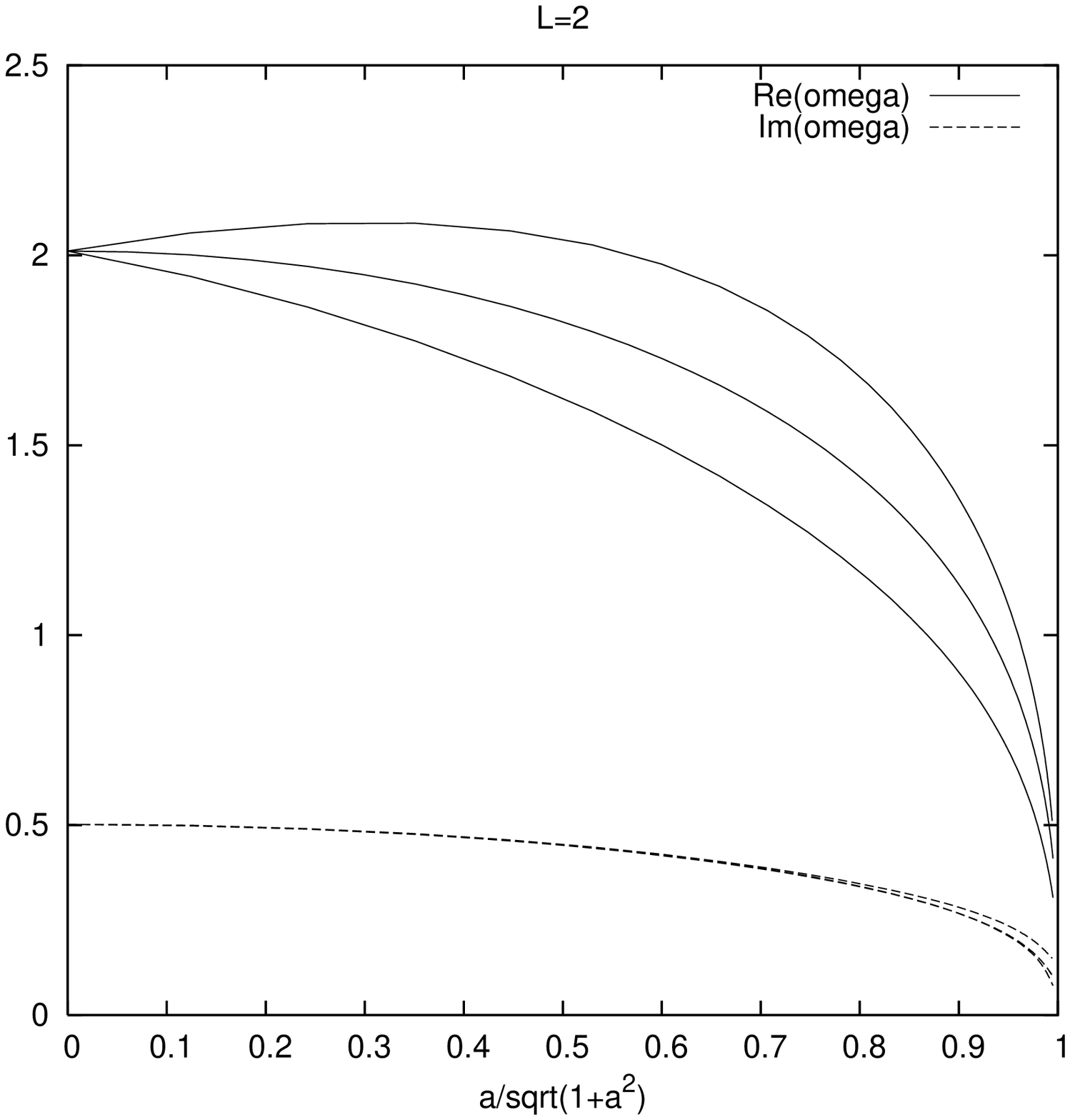}}}
\caption{\label{fig:L2}The quasinormal modes for $(j,m,\ell)=(0,\pm2,0)$
and $(0,0,1)$.}
\end{figure}

\begin{figure}
\rotatebox{270}{\resizebox{6cm}{!}{\includegraphics{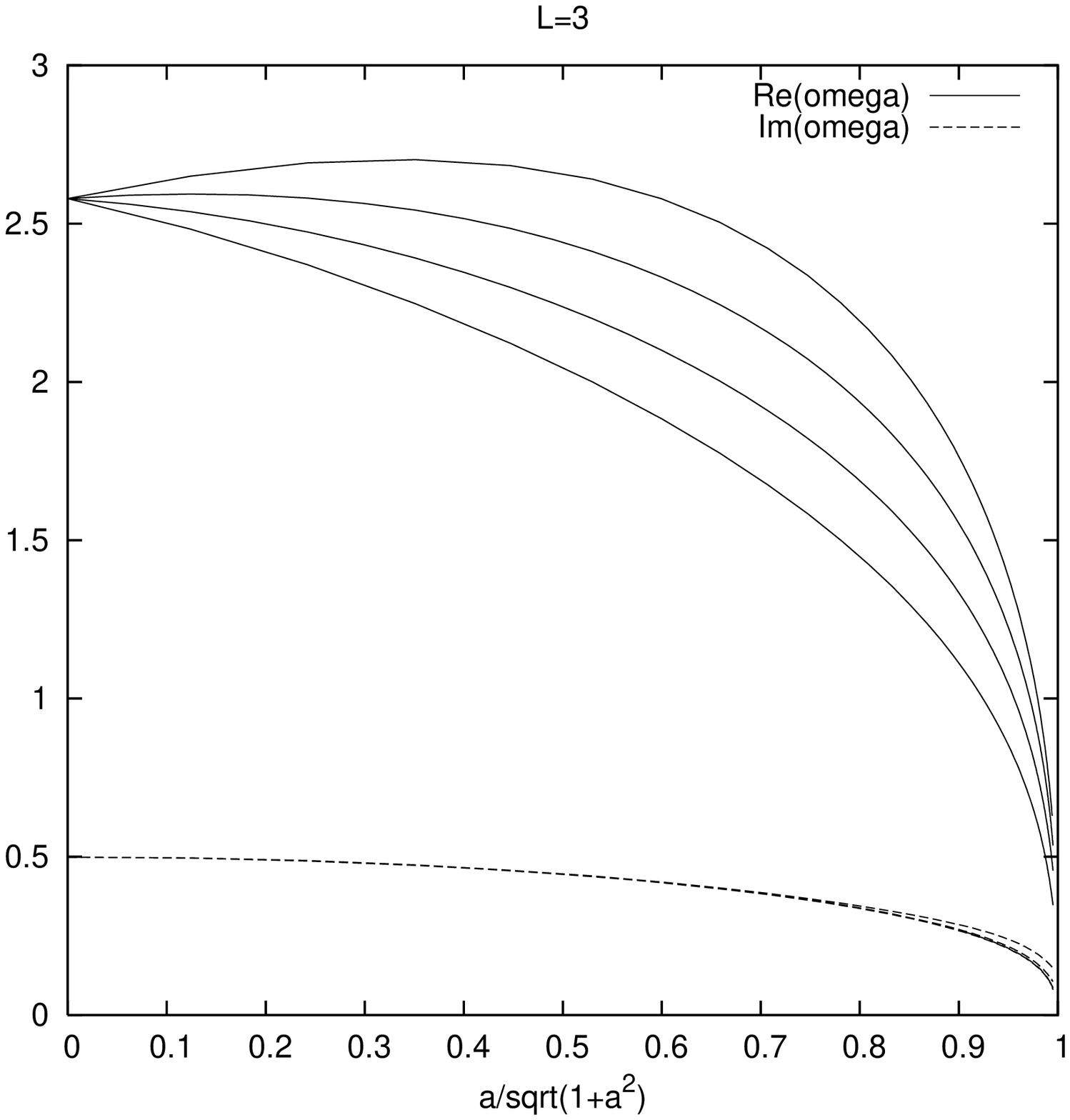}}}
\caption{\label{fig:L3}The quasinormal modes for $(j,m,\ell)=(0,\pm3,0)$
and $(0,\pm1,1)$.}
\end{figure}

\begin{figure}
\rotatebox{270}{\resizebox{6cm}{!}{\includegraphics{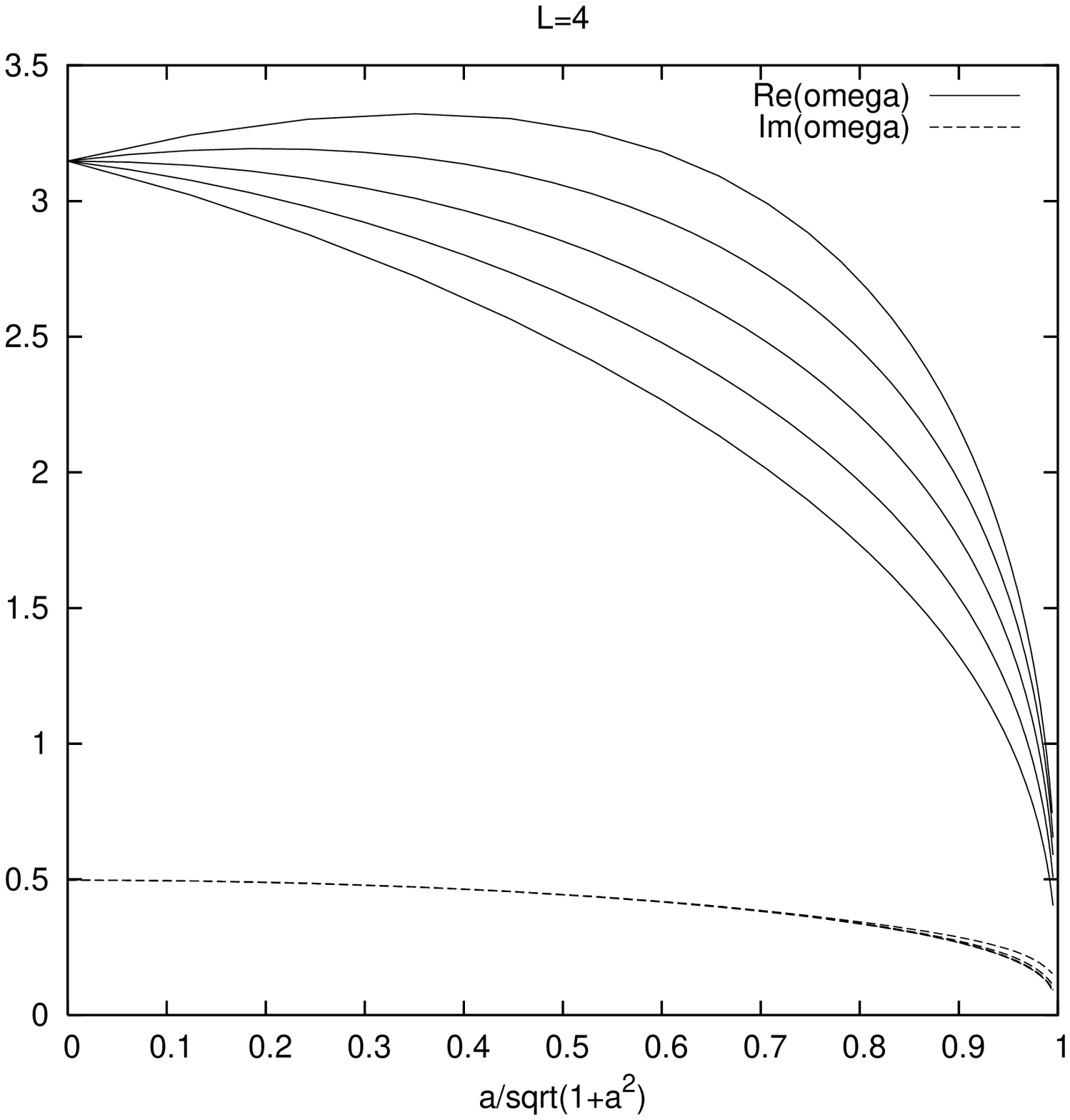}}}
\caption{\label{fig:L4}The quasinormal modes for $(j,m,\ell)=(0,\pm4,0)$,
$(0,\pm2,1)$ and $(0,0,2)$.}
\end{figure}

\begin{figure}
\rotatebox{270}{\resizebox{6cm}{!}{\includegraphics{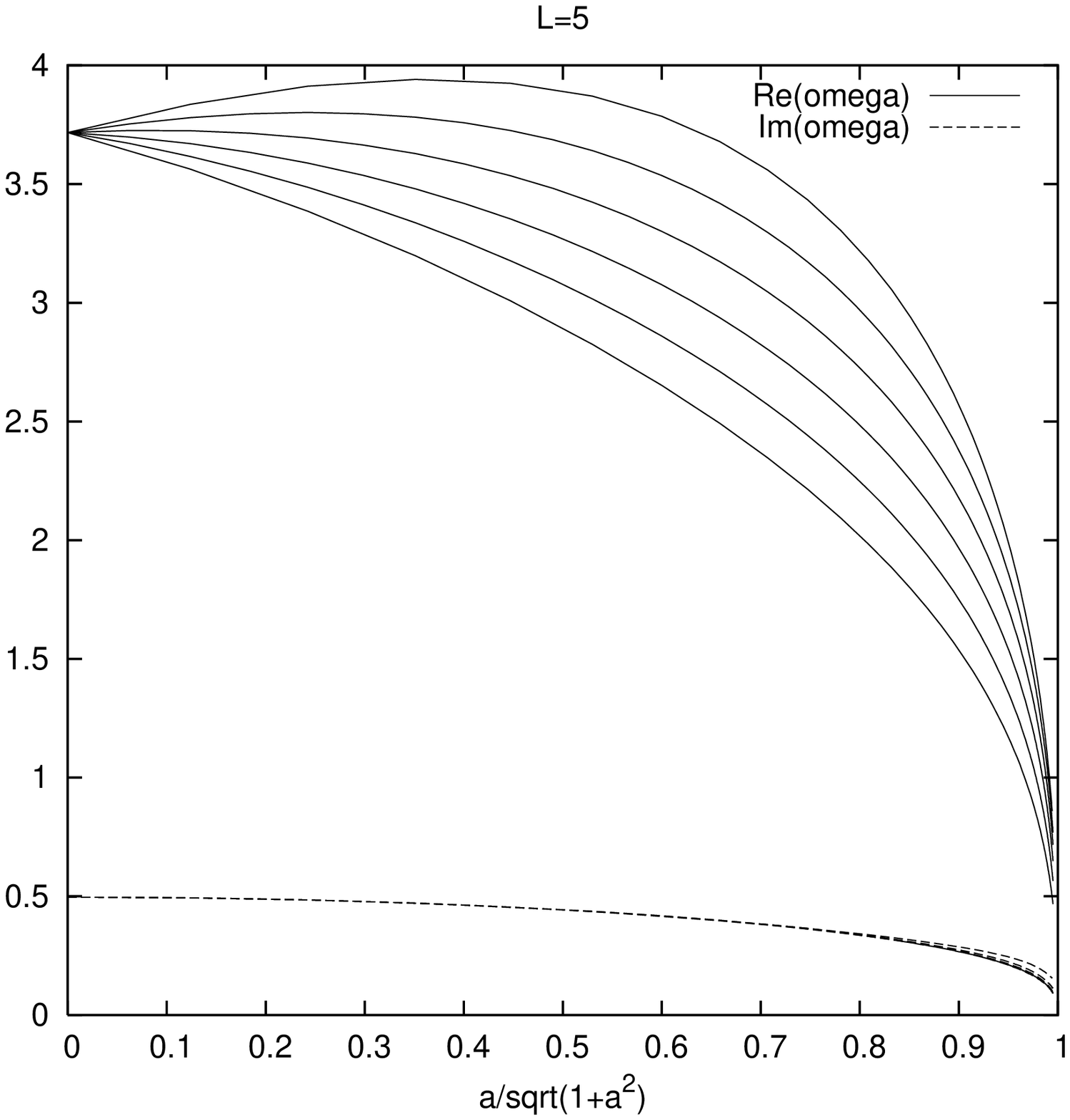}}}
\caption{\label{fig:L5}The quasinormal modes for $(j,m,\ell)=(0,\pm5,0)$,
$(0,\pm3,1)$ and $(0,\pm1,2)$.}
\end{figure}

Typical examples of quasinormal modes of the six-dimensional
Myers-Perry black holes are plotted in
Figs.~\ref{fig:L0}--~\ref{fig:L5}.
Here the real and imaginary parts of the resonant frequencies
$\omega_*$ as functions of the parameter $a_*/\sqrt{1+a_*^2}$ for
various value of $m$ and $\ell$ are plotted.
Although the Kerr parameter $a_*$ is taken to have a sufficiently
large value, each mode does not seem to have negative imaginary part
in any case.

\section{Summary and Discussion}
\label{sec:summary}
We investigated the massless scalar field equation in the
six-dimensional Myers-Perry black hole background.
Under the quasinormal boundary condition, we have searched for the
resonant modes.

In the non-rotating case ($a_*=0$), the quasinormal modes with same
$L(=j+|m|+2\ell)$ are degenerate.
Continuously varying the rotational parameter $a_*$, these modes split
into several modes characterized by $(j,m,\ell)$.
The sequences of quasinormal modes we obtained have positive imaginary
parts, thus they do not show any evidence for instability.

Recently, Cardoso, Siopsis and Yoshida~\cite{Cardoso:2004cj}
investigated the stability of the same system.
Several quasinormal modes within our present work are also calculated 
by them in different way to expand the radial function and their
results are consistent to our present results.
They also conclude that the stability of scalar field perturbations on 
Myers-Perry black holes are suggested.

\begin{acknowledgments}
The authors would like to thank G.~w.~Kang, and H.~Kodama for useful
discussions and comments.
Y.~M. is supported by a Grant-in-Aid for the 21st Century COE ``Center
for Diversity and Universality in Physics''.
\end{acknowledgments}

\appendix

\section{Gaussian eliminations}

The eight-term recurrence relations (\ref{eq:eight-term}) are reduced
to the seven-term ones
\begin{eqnarray}
\alpha'_kb_k+\beta'_kb_{k-1}+\gamma'_kb_{k-2}+\delta'_kb_{k-3}
+\epsilon'_kb_{k-4}+\zeta'_kb_{k-5}+\eta'_kb_{k-6}
&=& 0, \, (k=1,2,\cdots),
\end{eqnarray}
via the Gaussian elimination:
\begin{eqnarray}
&&
\alpha'_1=\alpha_1,\quad \beta'_1=\beta_1,
\nonumber\\&&
\alpha'_2=\alpha_2,\quad \beta'_2=\beta_2,\quad
\gamma'_2=\gamma_2,
\nonumber\\&&
\alpha'_3=\alpha_3,\quad \beta'_3=\beta_3,\quad
\gamma'_3=\gamma_3,\quad \delta'_3=\delta_3,
\nonumber\\&&
\alpha'_4=\alpha_4,\quad \beta'_4=\beta_4,\quad
\gamma'_4=\gamma_4,\quad \delta'_4=\delta_4,\quad
\epsilon'_4=\epsilon_4,
\nonumber\\&&
\alpha'_5=\alpha_5,\quad \beta'_5=\beta_5,\quad
\gamma'_5=\gamma_5,\quad \delta'_5=\delta_5,\quad
\epsilon'_5=\epsilon_5,\quad \zeta'_5=\zeta_5,
\nonumber\\&&
\alpha'_6=\alpha_6,\quad \beta'_6=\beta_6,\quad
\gamma'_6=\gamma_6,\quad \delta'_6=\delta_6,\quad
\epsilon'_6=\epsilon_6,\quad \zeta'_6=\zeta_6,\quad
\eta'_6=\eta_6,
\nonumber\\&&
\alpha'_k=\alpha_k,\quad
\beta'_k=\beta_k-\alpha'_{k-1}\theta_k/\eta'_{k-1},\quad
\gamma'_k=\gamma_k-\beta'_{k-1}\theta_k/\eta'_{k-1},\quad
\delta'_k=\delta_k-\gamma'_{k-1}\theta_k/\eta'_{k-1},\quad
\nonumber\\&&
\epsilon'_k=\epsilon_k-\delta'_{k-1}\theta_k/\eta'_{k-1},\quad
\zeta'_k=\zeta_k-\epsilon'_{k-1}\theta_k/\eta'_{k-1},\quad
\eta'_k=\eta_k-\zeta'_{k-1}\theta_k/\eta'_{k-1},
\quad (k=7,8,\cdots).
\end{eqnarray}
The similar procedures
\begin{eqnarray}
&&
\alpha''_1=\alpha'_1,\quad \beta''_1=\beta'_1,
\nonumber\\&&
\alpha''_2=\alpha'_2,\quad \beta''_2=\beta'_2,\quad
\gamma''_2=\gamma'_2,
\nonumber\\&&
\alpha''_3=\alpha'_3,\quad \beta''_3=\beta'_3,\quad
\gamma''_3=\gamma'_3,\quad \delta''_3=\delta'_3,
\nonumber\\&&
\alpha''_4=\alpha'_4,\quad \beta''_4=\beta'_4,\quad
\gamma''_4=\gamma'_4,\quad \delta''_4=\delta'_4,\quad
\epsilon''_4=\epsilon'_4,
\nonumber\\&&
\alpha''_5=\alpha'_5,\quad \beta''_5=\beta'_5,\quad
\gamma''_5=\gamma'_5,\quad \delta''_5=\delta'_5,\quad
\epsilon''_5=\epsilon'_5,\quad \zeta''_5=\zeta'_5,
\nonumber\\&&
\alpha''_k=\alpha'_k,\quad
\beta''_k=\beta'_k-\alpha''_{k-1}\eta'_k/\zeta''_{k-1},\quad
\gamma''_k=\gamma'_k-\beta''_{k-1}\eta'_k/\zeta''_{k-1},\quad
\delta''_k=\delta'_k-\gamma''_{k-1}\eta'_k/\zeta''_{k-1},\quad
\nonumber\\&&
\epsilon''_k=\epsilon'_k-\delta''_{k-1}\eta'_k/\zeta''_{k-1},\quad
\zeta''_k=\zeta'_k-\epsilon''_{k-1}\eta'_k/\zeta''_{k-1},\quad
\quad (k=6,7,\cdots),
\end{eqnarray}
\begin{eqnarray}
&&
\alpha'''_1=\alpha''_1,\quad \beta'''_1=\beta''_1,
\nonumber\\&&
\alpha'''_2=\alpha''_2,\quad \beta'''_2=\beta''_2,\quad
\gamma'''_2=\gamma''_2,
\nonumber\\&&
\alpha'''_3=\alpha''_3,\quad \beta'''_3=\beta''_3,\quad
\gamma'''_3=\gamma''_3,\quad \delta'''_3=\delta''_3,
\nonumber\\&&
\alpha'''_4=\alpha''_4,\quad \beta'''_4=\beta''_4,\quad
\gamma'''_4=\gamma''_4,\quad \delta'''_4=\delta''_4,\quad
\epsilon'''_4=\epsilon''_4,
\nonumber\\&&
\alpha'''_k=\alpha''_k,\quad
\beta'''_k=\beta''_k-\alpha'''_{k-1}\zeta''_k/\epsilon'''_{k-1},\quad
\gamma'''_k=\gamma''_k-\beta'''_{k-1}\zeta''_k/\epsilon'''_{k-1},\quad
\delta'''_k=\delta''_k-\gamma'''_{k-1}\zeta''_k/\epsilon'''_{k-1},\quad
\nonumber\\&&
\epsilon'''_k=\epsilon''_k-\delta'''_{k-1}\zeta''_k/\epsilon'''_{k-1},
\quad (k=5,6,\cdots),
\end{eqnarray}
\begin{eqnarray}
&&
\alpha''''_1=\alpha'''_1,\quad \beta''''_1=\beta'''_1,
\nonumber\\&&
\alpha''''_2=\alpha'''_2,\quad \beta''''_2=\beta'''_2,\quad
\gamma''''_2=\gamma'''_2,
\nonumber\\&&
\alpha''''_3=\alpha'''_3,\quad \beta''''_3=\beta'''_3,\quad
\gamma''''_3=\gamma'''_3,\quad \delta''''_3=\delta'''_3,
\nonumber\\&&
\alpha''''_k=\alpha'''_k,\quad
\beta''''_k=\beta'''_k-\alpha''''_{k-1}\epsilon'''_k/\delta''''_{k-1},\quad
\gamma''''_k=\gamma'''_k-\beta''''_{k-1}\epsilon'''_k/\delta''''_{k-1},\quad
\nonumber\\&&
\delta''''_k=\delta'''_k-\gamma''''_{k-1}\epsilon'''_k/\delta''''_{k-1},
\quad (k=4,5,\cdots),
\end{eqnarray}
\begin{eqnarray}
&&
\alpha'''''_1=\alpha''''_1,\quad \beta'''''_1=\beta''''_1,
\nonumber\\&&
\alpha'''''_2=\alpha''''_2,\quad \beta'''''_2=\beta''''_2,\quad
\gamma'''''_2=\gamma''''_2,
\nonumber\\&&
\alpha'''''_k=\alpha''''_k,\quad
\beta'''''_k=\beta''''_k-\alpha'''''_{k-1}\delta''''_k/\gamma'''''_{k-1},\quad
\gamma'''''_k=\gamma''''_k-\beta'''''_{k-1}\delta''''_k/\gamma'''''_{k-1},
\quad (k=3,4,\cdots),
\end{eqnarray}
leads to the three-term relations
\begin{eqnarray}
\alpha'''''_kb_k+\beta'''''_kb_{k-1}+\gamma'''''_kb_{k-2}=0,
\quad (k=1,2,\cdots).
\end{eqnarray}

\end{document}